\documentclass[runningheads]{llncs}
\usepackage{makeidx} 
\usepackage{graphicx}

\usepackage{booktabs}

\usepackage{hyphenat}

\usepackage{multicol}
\usepackage{multirow}

\usepackage{subcaption}

\usepackage{listings}

\usepackage[english]{babel}

\usepackage[svgnames]{xcolor}

\lstdefinelanguage{XES}
{
  basicstyle=\ttfamily\footnotesize,
  morestring=[b]",
  moredelim=[s][\bfseries\color{red}]{<}{\ },
  moredelim=[s][\bfseries\color{red}]{</}{>},
  moredelim=[l][\bfseries\color{red}]{/>},
  moredelim=[l][\bfseries\color{red}]{>},
  morecomment=[s]{<?}{?>},
  morecomment=[s]{<!--}{-->},
  commentstyle=\color{Gray},
  stringstyle=\color{Green},
  identifierstyle=\color{Maroon}
}

\begin{document}

\title{Extending predictive process monitoring\\ for collaborative processes}
\titlerunning{Extending predictive process monitoring for collaborative processes} 

\author{Daniel Calegari\inst{1,2}
\and
Andrea Delgado\inst{1}
}
\authorrunning{D. Calegari and A. Delgado}  

\institute{Instituto de Computación, Facultad de Ingeniería,Universidad de la República\\ 
\email{\{adelgado,dcalegar\}@fing.edu.uy}
\and
Universidad ORT Uruguay\\ 
\email{calegari@ort.edu.uy}
}

\maketitle

\begin{abstract}
Process mining on business process execution data has focused primarily on orchestration-type processes performed in a single organization (intra-organizational). Collaborative (inter-organizational) processes, unlike those of orchestration type, expand several organizations (for example, in e-Government), adding complexity and various challenges both for their implementation and for their discovery, prediction, and analysis of their execution. Predictive process monitoring is based on exploiting execution data from past instances to predict the execution of current cases. It is possible to make predictions on the next activity and remaining time, among others, to anticipate possible deviations, violations, and delays in the processes to take preventive measures (e.g., re-allocation of resources). In this work, we propose an extension for collaborative processes of traditional process prediction, considering particularities of this type of process, which add information of interest in this context, for example, the next activity of which participant or the following message to be exchanged between two participants.

\end{abstract}

\section{Introduction} \label{sec:introduction}

Business processes are coordinated sets of activities designed to achieve specific business objectives \cite{Weske19}. Their execution generates a wealth of data for process evaluation and continuous improvement, whether in traditional or process-oriented information systems. Process Mining techniques \cite{ProcessMiningBook}—encompassing discovery, conformance, and enhancement—enable complex, in-depth analyses of actual process executions. These techniques provide organizations with crucial insights into efficiency, quality, and regulatory compliance, thereby uncovering opportunities for evidence-based improvements. Such post-mortem analyses are invaluable, forming the foundation for systematic process enhancement and supporting informed decision-making.

Nevertheless, organizations demand the ability to exploit historical execution data and real-time observations to forecast the future states or outcomes of ongoing business process instances. Predictive Process Monitoring \cite{DiFrancescomarino2022} is a subfield of Process Mining that focuses on this aspect. Using event logs as input makes it possible to make accurate predictions regarding process execution, such as anticipating potential deviations, violations, or delays. These predictions enable proactive interventions—such as resource reallocation or adjustments in time management—to mitigate risks and ensure smooth process execution.

Traditionally, Process Mining research has concentrated on orchestration-type processes contained within a single organization (intra-organizational). In contrast, collaborative processes (inter-organizational), which involve multiple organizations (e.g., in e-Government), present additional complexity. Due to their inherently distributed and collaborative nature, these processes introduce challenges in implementation, discovery, prediction, and analysis \cite{Aalst11,PenaA0C23}.

Considering their unique characteristics, addressing the prediction needs for collaborative processes is critically important for organizations. For instance, beyond merely applying and evaluating existing ``as-is'' techniques for predicting the remaining execution time of a trace or the next event, it is essential to integrate or extend these methods. Incorporating elements such as the execution history of the involved participants can enable more accurate predictions of interest tailored to the collaborative context.

The primary objective of this work is to analyze existing techniques and approaches for predicting the execution of inter-organizational collaborative business processes using Process Mining. Moreover, it seeks to define, extend, or adapt predictive methods for collaborative business processes, emphasizing e-Government environments. This research aims to bridge the gap in predictive analytics for collaborative processes, providing organizations with tools to better manage and optimize their cross-organizational workflows.

The rest of the paper is structured as follows. 
We discuss related work in Section \ref{sec:relatedwork}.
In Section \ref{sec:eventlogs}, we present collaboration event logs used as input for predictions.
In Section \ref{sec:proposal}, we present how prediction can be conceived in the context of collaborative processes, and in Section \ref{sec:evaluation}, we present an example application.  
Finally, in Section \ref{sec:conclusions}, we conclude and outline future work.

\section{Related work} \label{sec:relatedwork}

In recent years, different approaches to predictive process monitoring have been applied and evaluated \cite{DiFrancescomarino2022,Marquez-Chamorro18}. These approaches use different encodings of the problem, exploiting an explicit representation of the process model to make the prediction or not, and also addressing classification or regression problems based on the type of predicted value (categorical or numerical). 
In this context, works are allowing remaining time (e.g., \cite{AalstSS11}), outcome (e.g., \cite{Hong16}), next event (e.g., \cite{Breuker15}), and deviation (e.g., \cite{GrohsPR23}) predictions, among others, that can be applied to business processes, as well as other types of industrial processes such as production, manufacturing or case-handling processes. More recently, works have spread into analyzing object-centric event logs, e.g., \cite{GalantiLNM23,SmitRL24}.

Most of the predictive methods are focused on the predicted value of an individual process instance \cite{DiFrancescomarino2022,Marquez-Chamorro18}. Few works propose predictive process monitoring in the context of multiple process instances. It poses new challenges since predictions heavily depend on other cases running simultaneously, e.g. when cases share limited resources. 
In \cite{CONFORTI20151}, the authors describe a recommendation system determining the probability of a risk in a system. It deals with the interplay between risks relative to multiple process instances running concurrently. 
As another example, in \cite{SENDEROVICH2019255}, the authors propose a method for feature encoding within a bi-dimensional space characterized by intra- and inter-case features.

Although these works considered multiple process instances, they did not consider such instances within a collaborative environment, as in the case of collaborative business processes. Few works consider this collaborative context. 
In \cite{Cao23}, the authors propose a strategy for predicting remaining time in collaborative business processes while preserving organizations' privacy, i.e., by sharing hidden state information between sub-models. As far as we know, this is the only work that analyzes predictive process monitoring for collaborative processes.
\section{Event logs for collaborations} \label{sec:eventlogs}

Making process predictions requires training, in which a predictive model is learned from historical (complete) execution traces in the form of an event log. Such models are then queried to predict the future of an ongoing case \cite{DiFrancescomarino2022}.

Within a collaborative environment, several organizations (or participants of an organization) interact with each other to carry out a global process. Single orchestrations exist for each participant, and messages allow them to coordinate.

Traditional event logs only sometimes register data regarding which participant enacts the activity of the corresponding role/person. To cope with this limitation, we have defined an extension \cite{GonzalezD21CLEI} of the XES format for collaborative business process, which was tested for discovering collaborative processes \cite{PenaA0C23}. A collaboration event log comprises collaborative cases involving several participants whose events come from different participants and include mandatory attributes: \texttt{participant} to identify the participant that enacts the event, \texttt{elemType} with the type of the event (user, message), and if the event is of type \texttt{message}, from which participant it is being received (\texttt{fromParticipant}) or to which participant it is being sent (\texttt{toParticipant}). As described in \cite{PenaA0C23}, a collaboration event log can be built from the participants' logs by merging them.

\begin{figure}[!ht]
     \centering
     \begin{subfigure}[a]{\textwidth}
         \centering
         \includegraphics[width=0.8\linewidth]{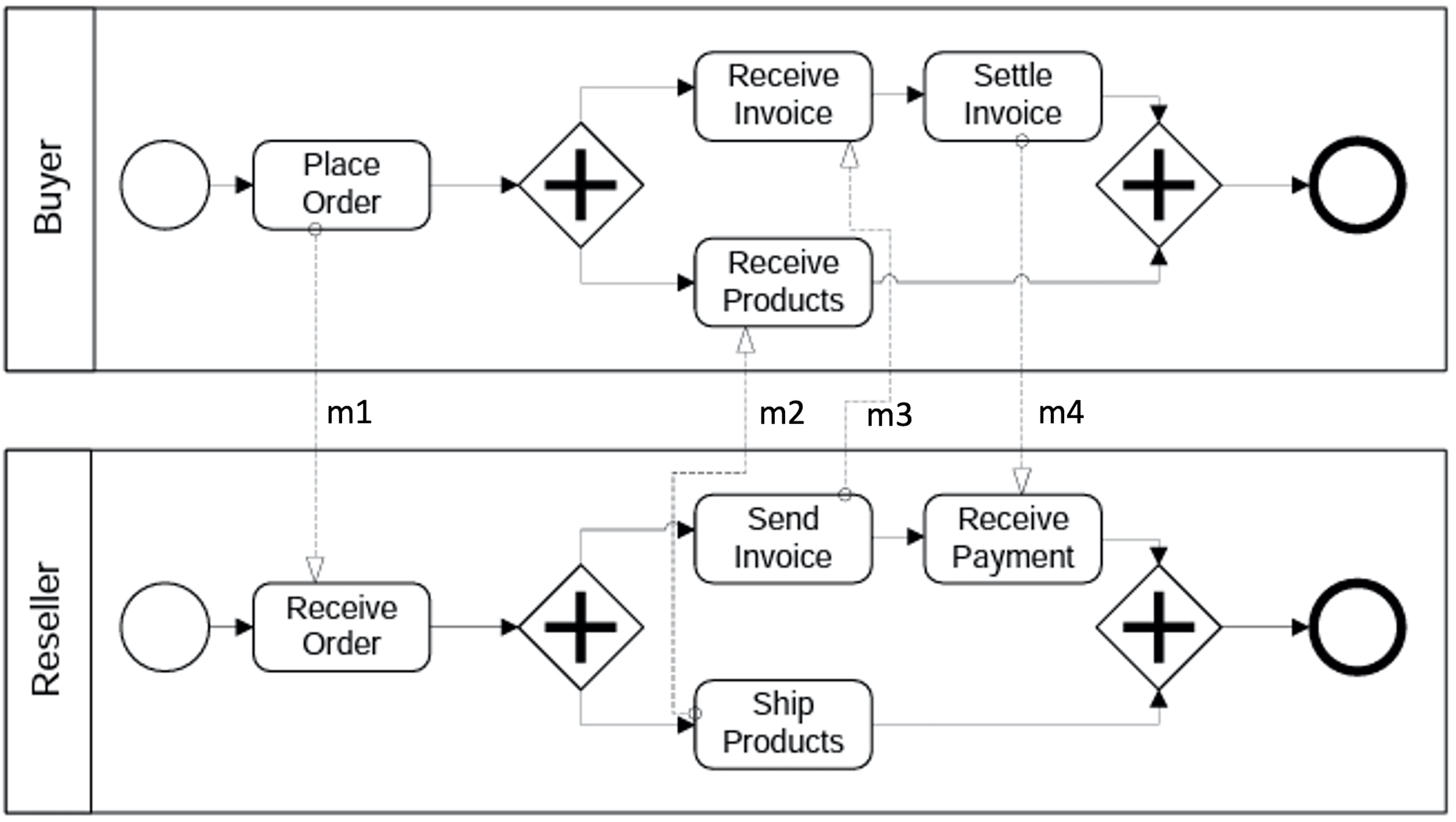}
    \caption{Example collaborative business process with messages from \cite{Weske19} }
    \label{fig:ExampleCollabWeskeprocess}
    \end{subfigure}
    \hfill
    \begin{subfigure}[t]{0.4\textwidth}
         \centering
         \includegraphics[width=\textwidth]{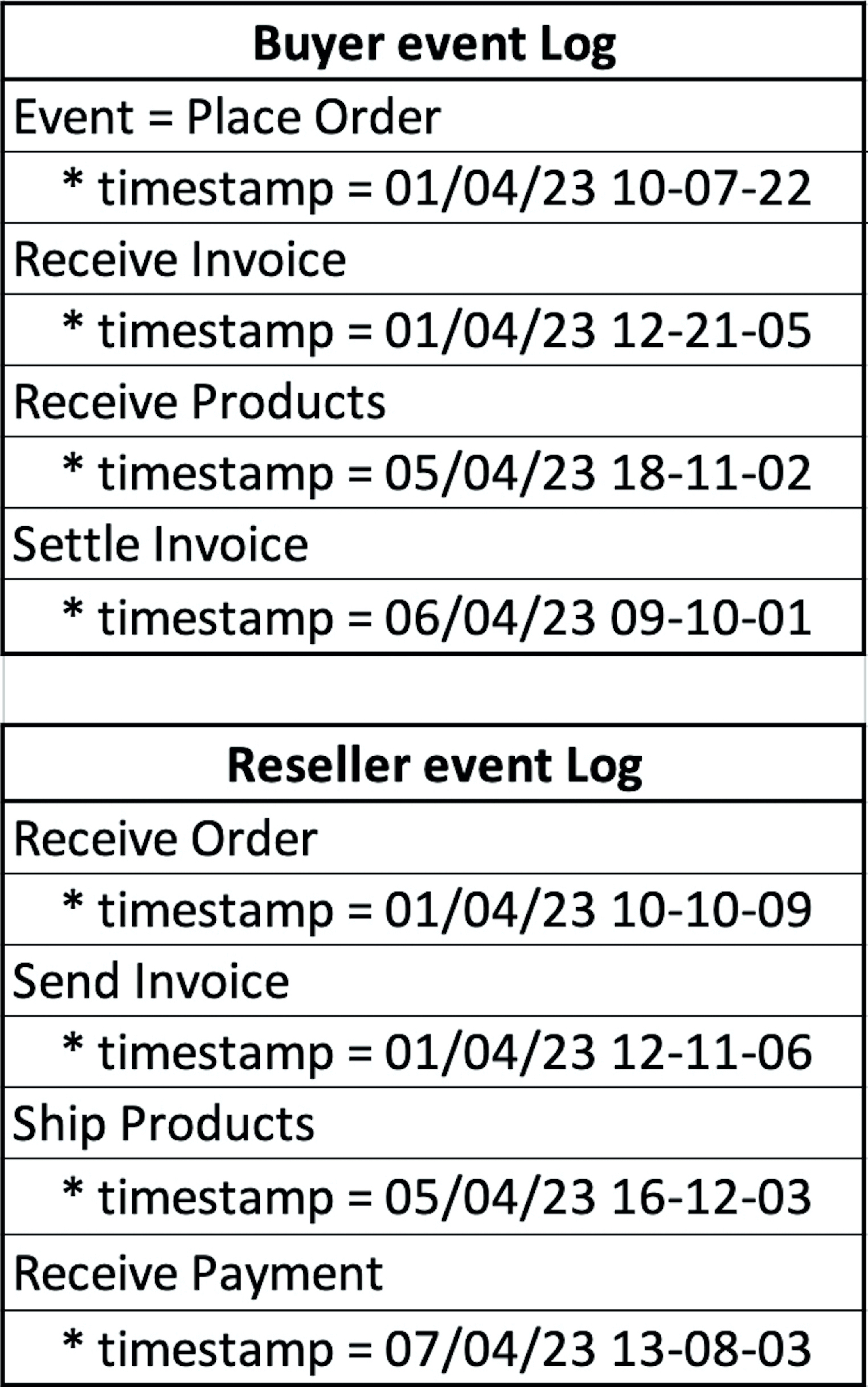}
         \caption{Participants' reduced event logs (orchestration)}
         \label{fig:ExampleLogsCollabWeske_A}
     \end{subfigure}    
     \begin{subfigure}[t]{0.4\textwidth}
         \centering
         \includegraphics[width=\textwidth]{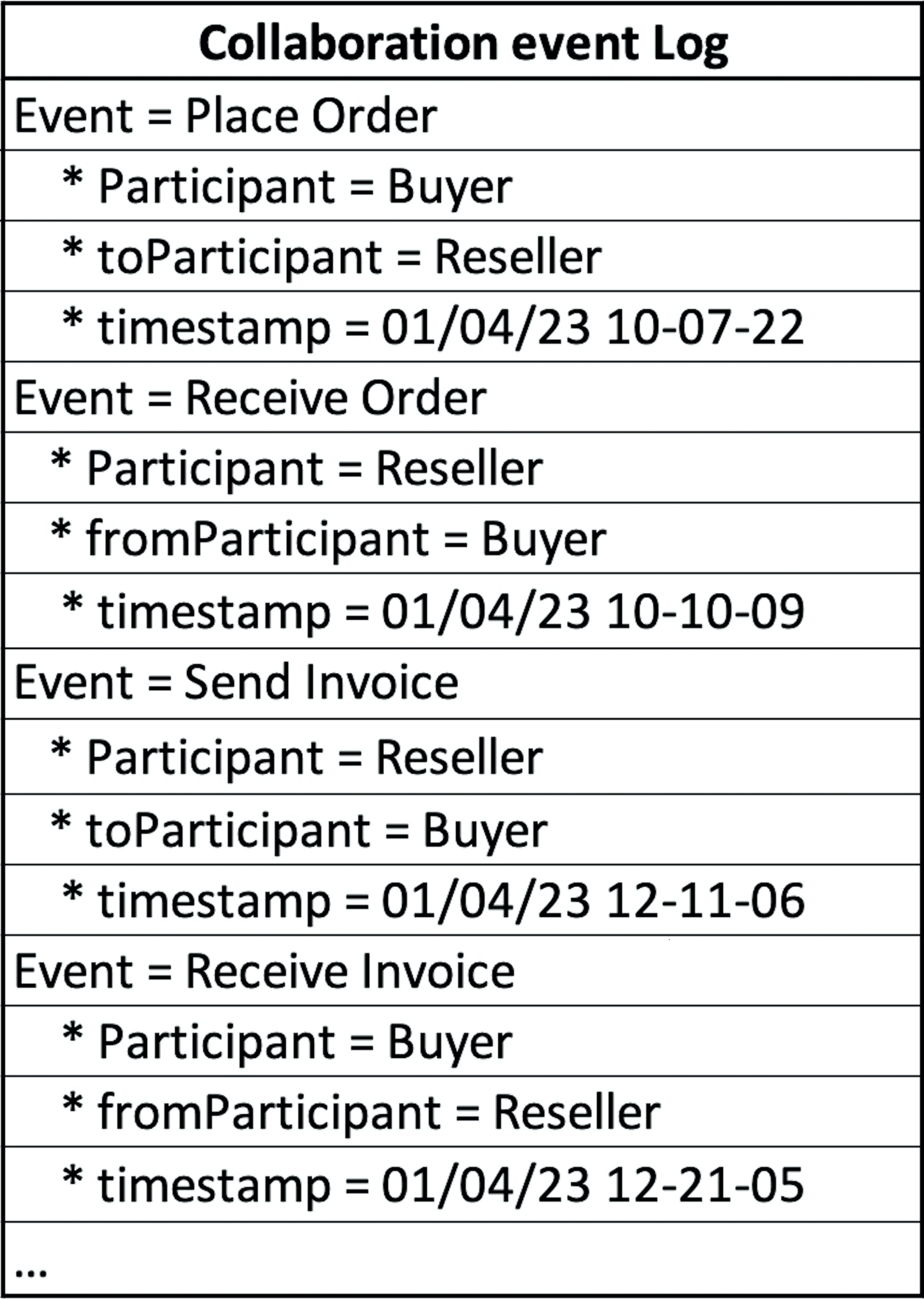}
         \caption{Extended event log for collaboration}
         \label{fig:ExampleLogsCollabWeske_B}
     \end{subfigure}    
    \caption{Example collaborative process and associated event logs from  \cite{PenaA0C23}}
    \label{fig:ExampleCollabWeske}
\end{figure}

In Figure \ref{fig:ExampleCollabWeskeprocess}, a reseller company's ordering process is depicted as a collaborative business process between a Buyer and a Reseller. Figure \ref{fig:ExampleLogsCollabWeske_A} shows an example of an event log for each participant, and Figure \ref{fig:ExampleLogsCollabWeske_A} shows an extended event log for the collaboration. The collaboration event log merges each participant's event log, considering the timestamp to maintain the order of execution in the general view of the collaborative case. 

\section{Predictive collaborative process monitoring} \label{sec:proposal}

According to the literature, it is possible to classify the existing prediction types into three categories \cite{DiFrancescomarino2022}: 
\begin{enumerate}
    \item Outcome-based: predictions related to predefined categorical or boolean outcome values, e.g., predicting if a particular process path will occur.
    \item Numeric value: predictions related to measures of interest taking numeric or continuous values, e.g., time prediction such as estimating how long it will take for a currently running process instance to complete.
    \item Next event: predictions related to sequences of future activities and related data payloads, e.g., forecasting the next step or event likely to occur.
\end{enumerate}

Considering the importance of main collaborative concepts (participants and messages), the kind of predictions could be extended, as summarized in Table \ref{tab:predtypes}.

\begin{table}[thb]
\setlength{\tabcolsep}{3pt}
\caption{Non-exhaustive type of predictions for collaborative processes}%
\label{tab:predtypes}
\begin{tabular}
{l|l}
\toprule
\bf Category & \bf Predictions \\
\midrule
Outcome-based & 
If a participant will participate in a case. \\
& If a particular message will be sent/received.\\
\midrule
Numeric value & 
Number of remaining/total messages of a participant.\\
& Number of remaining/total messages in the process.\\
& Participant remaining/duration time (other could be active). \\
 & Process remaining/duration time (every participant finishes). \\
 & Time until the next message to send/receive. \\
\midrule
Next Event & Next event that is likely to occur in a participant. \\
 & Next event that is likely to occur in the process. \\
 & Next participant that is likely to act. \\
 & Next participant that is likely to send/receive a message. \\
 & Next message (send/receive) that is likely to occur in the process.\\ 
 & Next message (send/receive) that is likely to occur in a participant.\\
\bottomrule
\end{tabular}
\end{table}

Broadly speaking, the possibilities are divided between predictions linked to the process as a whole or a specific participant (a problem reduced to forecasts within a single organization). For example, concerning the model in Figure \ref{fig:ExampleCollabWeskeprocess}, it could be possible to focus predictions from the perspective of the Buyer and estimate what will be the next event if the case is on the place order activity or the remaining time of such a participant. If considering the process as a whole, the next activity can be within the Buyer or the Reseller (e.g., receiving an order), and the remaining time needs to consider both participants.
Moreover, it is possible to identify messages as events that occur and are registered, similar to process activities. Again, in the Buyer's example, it could be possible to estimate whether the following message to receive will be the receipt of the product (m2) or the invoice (m3).

\subsection{Adapting predictive methods}

As explained next, the problem of making predictions within a collaborative environment could be reduced to the problem of making predictions for single orchestrations. 
The critical point is the form of the collaboration event log itself. Since it merges each participant's event log, it is like an orchestration log. Moreover, for each event, it is possible to identify the participant who enacts it and whether the event concerns the sending/reception of a message. This information takes the form of attributes that can be used for outcome-based and numeric value predictions. 

Notice that many encodings and prediction methods are possible. For example, to predict the following message that is likely to occur within a participant, it is possible to take all the events of a case, only the events of a particular participant, or only the message events of the case or participant. Each one of these options offers more or less information concerning the problem to resolve, which is later used to train the predictive model. As a result, it could yield more or less precise results. This aspect must be evaluated in depth in future work.

\subsubsection{Outcome-based predictions} could be addressed by taking the participant or message names as the actual (categorical) values of a variable that we aim to predict, similar to any outcome-based prediction in the case of an orchestration. The encoding of traces could consider the whole process, e.g., in the case of predicting the participant, or a concrete participant, e.g., in the case of predicting a particular message.

\subsubsection{Numeric value predictions} such as the number of remaining/total messages can be addressed similarly to the number of remaining/total events in an orchestration. Event logs for training could consider the events or only messages of the whole process or a particular participant. The case of predicting a participant's remaining/duration time is precisely the orchestration case. To predict the same in the case of the whole process, we can take the entire event log as being of one participant. Finally, predicting the time until the following message is similar to predicting the time until the next event of a particular type.

\subsubsection{Next event predictions} are the most straightforward and only require taking the event log as a whole or reducing it for a given participant. The case of predicting the following message is similar to predicting the next event of a particular type (if any). Finally, predicting the next participant is similar to predicting the next event with the desired participant as an attribute.
\section{Assessment} \label{sec:evaluation}

We have performed a preliminary assessment of the ideas by implementing a prediction tool based on the work in \cite{bukhsh2021}. In such work, the authors propose ProcessTransformer, an approach for learning high-level representations from event logs with an attention-based network. The transformer allows for predicting the next activity, the event time, and the remaining time for a running case. We have used such a transformer as a basis and performed the transformations mentioned in the last section to make collaborative predictions.

\begin{figure}[!ht]
    \centering
    \includegraphics[width=\linewidth]{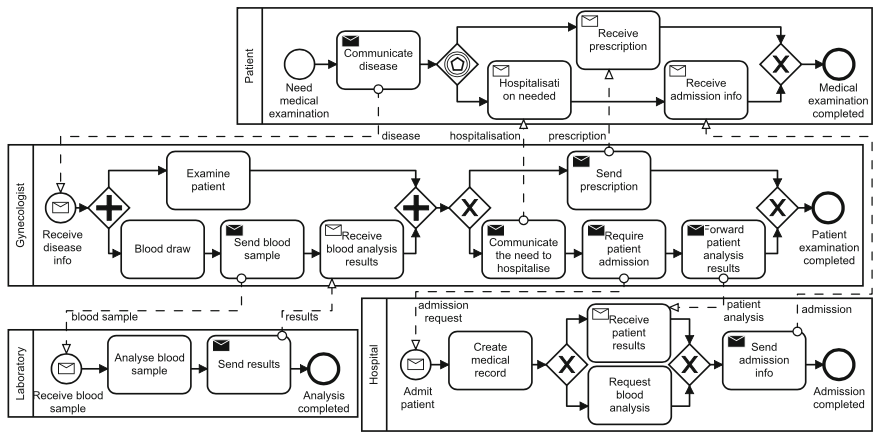}
    \caption{A healthcare business process collaboration from \cite{Lorenzo22} }
    \label{fig:healthcare}
\end{figure}

As an application example, we use the healthcare collaborative process introduced in \cite{Lorenzo22} and depicted in Figure \ref{fig:healthcare}. The process illustrates a healthcare scenario about the treatment of gynecological diseases.

\begin{figure}[!ht]
     \centering
     \begin{subfigure}[a]{\textwidth}
        \centering
        \includegraphics[width=0.8\linewidth]{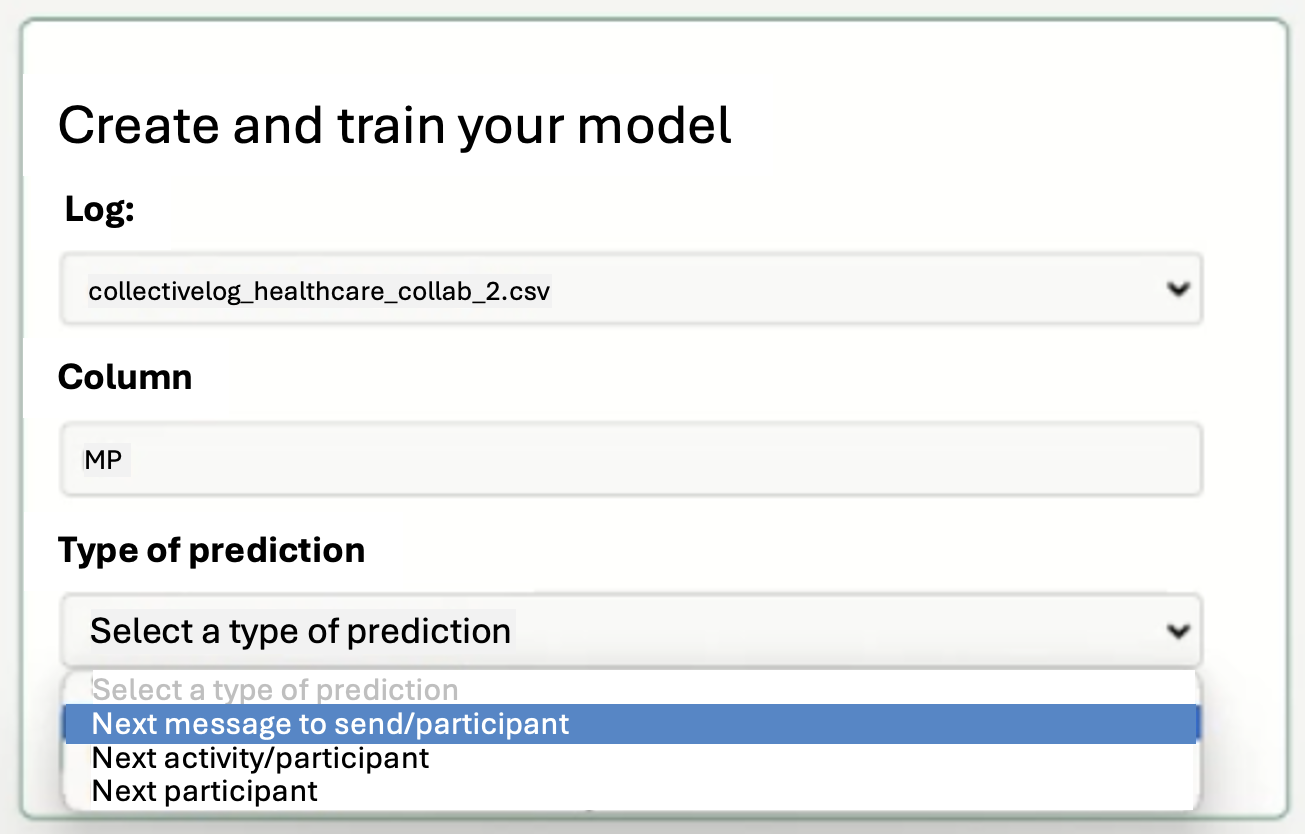}
    \caption{Training phase}
    \label{fig:tooltraining}
    \end{subfigure}
    \hfill
    \begin{subfigure}[t]{0.9\textwidth}
         \centering
         \includegraphics[width=\textwidth]{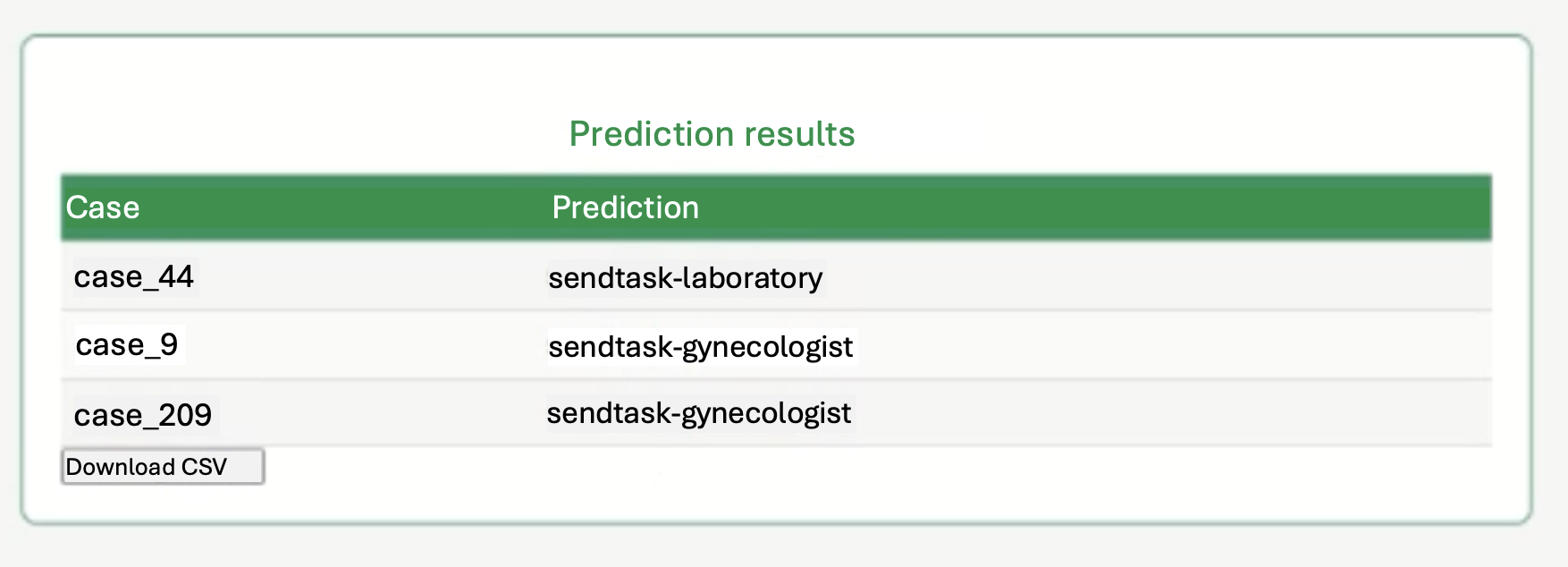}
         \caption{Prediction result}
         \label{fig:toolresult}
     \end{subfigure}    
    \caption{Web tool training and prediction results}
    \label{fig:tool}
\end{figure}

We took 
several collaborative logs conforming to such a process. The tool allowed the upload of a log with the complete traces to be used for the training phase. Figure \ref{fig:tooltraining} depicts how the log is uploaded and the type of prediction selected. In this case, we were interested in the following send message/participant that is likely to occur. We also selected the column that stores the information for training the model, i.e., the message information. Every other event, not being a message, was ignored for training (just for the example).

After training, a second page allows uploading a new log (of incomplete traces), and for each trace, the transformer predicts the following send message. There are three cases in the example depicted in Figure \ref{fig:toolresult}. The first case (case\_44) corresponds to a trace where the Patient participant already communicated the disease to the Gynecologist (start event Receive disease info), and the Gynecologist already sent the Send blood sample message to the LaboratoryLaboratory (from the second path in the parallel gateway) and its corresponding reception (Receive blood sample).  
Based on this information, the prediction correctly retrieves the message to be sent back to the Gynecologist (Send results). 

In the other two traces, predictions correspond to messages to be sent from the Gynecologist participant, e.g., the second trace (case\_9) corresponds to the Send blood sample message. All predictions can be downloaded as a .csv file. 
\section{Conclusions} \label{sec:conclusions}

In this work, we analyzed how traditional predictive process monitoring could be extended to inter-organizational collaborative business processes. Although concepts such as participant and message become more relevant in this context, conventional prediction techniques can be adapted to consider them.

We theoretically analyzed how this adaptation could be performed and then showed preliminary results of how some of the predictions could be achieved practically. Since our analysis is preliminary, other prediction types could be defined, and new techniques can be explored and tailored in this context. As mentioned, many possible encodings and prediction methods exist for the same problem. In this context, a more rigorous experimental evaluation is necessary to assess the quality of the predictions and the suitability of methods for predicting certain specific aspects of interest.

\section{Acknowledgment}
We want to thank the undergraduate students who worked on the practical application: Carolina Espino and Nicolás Ribero. Supported by project ``Minería de procesos y datos para la mejora de procesos colaborativos aplicada a e-Government'' funded by Agencia Nacional de Investigación e Innovación (ANII), Fondo María Viñas (FMV) "Proyecto ANII N° FMV\_1\_2021\_1\_167483", Uruguay

\bibliographystyle{splncs04}
\bibliography{bibliography.bib}

\end{document}